\documentclass[sigconf]{acmart}
\AtBeginDocument{%
  }

\setcopyright{acmlicensed}
\copyrightyear{2026}
\acmYear{2026}
\acmDOI{XXXXXXX.XXXXXXX}
\acmConference[CHI Workshop '26]{Make sure to enter the correct
  conference title from your rights confirmation email}{April 13--17,
  2026}{Barcelona, Spain}
\acmISBN{978-1-4503-XXXX-X/2018/06}

\usepackage{soul}

\begin{document}

\title{Who's Sense is This? Possibility for Impacting Human Insights in AI-assisted Sensemaking}

\author{Zhuoyi Cheng}
\orcid{0009-0003-3941-8869}
\affiliation{%
  \institution{Eindhoven University of Technology}
  \city{Eindhoven}
  \country{Netherlands}
}
\email{z.cheng@tue.nl}

\author{Steven Houben}
\orcid{0000-0002-9009-5706}
\affiliation{%
  \institution{Eindhoven University of Technology}
  \city{Eindhoven}
  \country{Netherlands}
}
\email{s.houben@tue.nl}

\renewcommand{\shortauthors}{Cheng and Houben}

\begin{abstract}
Sensemaking is an important preceding step for activities like consensus building and decision-making. When groups of people make sense of large amounts of information, their understanding gradually evolves from vague to clear. During this process when reaching a conclusion is still premature, if people are presented with others' insights, they may be directed to focus on that specific perspective without adequate verification. We argue that similar phenomena may also exist in AI-assisted sensemaking, in which AI will usually be the one that presents insight prematurely when users' understandings are still vague and ill-formed. In this paper, we raised three questions that are worth deliberation before exploiting AI to assist in collaborative sensemaking in practice, and discussed possible reasons that may lead users to opt for insights from AI.
\end{abstract}





\maketitle

\section{Introduction}
Sensemaking is an important activity for groups to understand information and hence support follow-up actions, like consensus building and decision-making~\cite{Liu2018ConsensUs}. But making sense of large amounts of information is never trivial work. Previous practices facilitating individuals and groups in sensemaking task primarily help people organize and visualize information by creating structured externalization, like tables~\cite{Chang2020Mesh} and diagrams~\cite{Subramonyam2020texSketch}. These externalizations are clear re-presentations of existing information, but creating which risks burdening users with transformed workflow and prematurely formalized input.

Now that AI is capable of comprehending large amounts of data and generating multi-modal outputs, researchers have started to explore its potential to address previous problems in sensemaking and even transform how people make sense of information. However, AI's handiness are double-edged. As sensemaking is a process for human understanding to gradually forage, vagueness exists most of the time. On the one hand, users' cognitive overhead may be lowered thanks to the possibility of vague expression (e.g., scattered sentences) when clear insights have not been formed in the early stage, compared to input methods like assigning values on sliders. On the other hand, current AI tends to disambiguate such vagueness and articulate insights via Chain-of-Thought (CoT), in which synthesis and re-interpretation of original information happens~\cite{Kang2023Synergi, Liu2024Selenite}, and hypotheses not originated from humans may emerge and be leveraged in even premature stages. Researchers have found that, in human-human collaborative sensemaking, users may risk focusing on specific aspects without adequately analyzing the quality of the insight when presented with certain insights in the premature stage~\cite{Goyal2016Translucence}. We argue that this could also happen in AI-assisted sensemaking and thus impacting human insights and cause a series of problems.



\section{Potential Problem}
To understand the potential problems of human insights being impacted, let's consider two scenarios that often require people to make sense of complex and abundant information: crime analysis~\cite{Goyal2016Translucence} and travel planning~\cite{Hargreaves2010Planning}.

Crime analysis is a high-stakes, high-responsibility context where unbiased and objective reasoning matters. If AI helps investigators make sense of evidence, there are chances that AI can chain clues and rationalize logic between events by making assumptions even if the chain of evidence is incomplete. Moreover, if such insights are presented inappropriately, for instance, diagrams with textual explanation in an assertive tone, it can more or less affect investigators' judgment and way of relating clues, leading to biased results. For crime analysis and contexts alike, biased sensemaking results could have fatal impacts for later decision-making and ultimately social safety and justice.

In contrast, travel plan is an activity that may have much less impact on the society and value humans' subjective preferences instead of objective correctness. Nevertheless, considering the convenience of AI-assisted sensemaking, we worry that humans' subjective preferences may be overshadowed, affected, or even ``manipulated'' implicitly, especially when AI service providers are also involved in marketing domains. For example, when selecting accommodations based on a set of travel constraints, is it possible that brands sponsored by the AI company will gain higher chance to be recommended to the users? In this case, AI-assisted travel information sensemaking could involve unfair business practices. In conclusion, in real sensemaking practices, if the insights are primarily contributed by AI, we believe that later consensus building and decision-making processes that include interpersonal interactions may risk being guided by AI instead of humans, which urges us to raise the following questions:


\begin{enumerate}
    \item Will perspectives from people be affected, diluted, or even excluded from the sensemaking process if insights from AI are presented prematurely? If so, what are the impacts?
    \item How would the follow-up consensus building and decision-making activities be affected if:
    \begin{enumerate}
        \item Adopted AI insights are misleading, biased, or incorrect;
        \item Adopted AI insights are logically correct, but humans' subjective preferences matter.
    \end{enumerate}
    \item How can we alleviate the above problems through proper designs of the interaction, interface, and AI behaviors?
\end{enumerate}



\section{Possible Reasons for Human Insights Being Impacted}
In this section, we discussed three possible reasons for users to tend to adopt AI insights in AI-assisted sensemaking, which may ultimately result in their own perspectives being affected, diluted, or even excluded.

\subsection{Algorithm Appreciation}
In some cases, people are more willing to opt to AI sensemaking results than other people and even themselves, which is called \textit{algorithm appreciation}~\cite{Logg2019appreciation}. However, whether the users will appreciate algorithmic results, and the extent of users' algorithm appreciation differ according to the final goal, the consequences of grounding follow-up on inappropriate insights, and user demographics. By comparing the studies of~\citet{Dietvorst2015aversion} and~\citet{Logg2019appreciation}, we infer that when faced with situations where negative consequences may occur if decisions are not wisely made, like university admission, people tend to have algorithm aversion rather than appreciation. In contrast, tasks like predicting popular songs will have trivial impact even if the prediction is wrong, and people tend to have algorithm appreciation in such situations no matter how much subjective preference weighs. Besides, lay people who are neither less numerate nor domain expert tend to choose algorithm results more~\cite{Logg2019appreciation}. As sensemaking per se covers a wide landscape and is closely related to consensus building and decision-making, which further enlarge the landscape, the occurrence of algorithm appreciation and adoption of AI insights is almost inevitable.


\subsection{Over-reliance on AI}
Although over-reliance on AI has not been extensively discussed in sensemaking tasks, we found previous research in design explicitly identified and discussed this problem~\cite{Chen2025CoExploreDS, Wadinambiarachchi2024DesignFixation}, and we believe that the findings are applicable to sensemaking because both tasks feature open-endedness that asks users to select between alternatives to proceed to the next step. Drawing on research in AI-assisted design, we conclude that users may over-rely on AI because AI generates satisfactory results in early stages that let users form a positive image towards AI~\cite{Chen2025CoExploreDS, Nourani2021Anchoring}. Besides, AI is superior to humans in output efficiency: it can read through information and give structured output that could take humans tens of minutes in a few seconds, and even recommend plausible next steps and directions at the same time. This convenience and huge efficiency enhancement could tempt people to take advantage of what AI generates directly, and gradually rely more and more often on AI~\cite{Wadinambiarachchi2024DesignFixation}.



\subsection{Implicit Persuasion}
Technologies can be persuasive, and the more persuasive they are, the greater the chance for users to adopt AI insights in sensemaking tasks. Humans have an intention to treat computers as social actors, through which computers can implicitly affect human behaviors and attitudes through social cues~\cite{Fogg2002Persuasive}. As current AI is getting increasingly anthropomorphic than previous technologies, its ability to persuade humans is getting stronger. For example, simple pre-scripted notifications with emotions like ``You've got mail!'' can lead people to infer that the system might be animate, whereas generative AI exhibits higher understanding of social dynamics and could talk in contextualized and personalized ways, which could strengthen AI's persuasive social role. In addition to the intrinsic social features AI has, we argue that how AI insights are presented may also lead to different levels of persuasiveness. For example, graphical presentations have been demonstrated to affect human attitudes more than textual ones~\cite{Fogg2002Persuasive}, and even if the modal remains the same, text for example, could have different effects when characteristics like tone and format differ. 




\begin{acks}
This publication is part of the project CurateAI with file number VI.Vidi.233.114 of the research programme NWO VIDI which is (partly) financed by the Dutch Research Council (NWO) under the grant \url{https://doi.org/10.61686/EPRLP31224}.
\end{acks}

\bibliographystyle{ACM-Reference-Format}
\bibliography{sample-base}

\end{document}